\documentclass[pdflatex,sn-mathphys-num]{sn-jnl}

\usepackage{graphicx}
\usepackage{multirow}
\usepackage{amsmath,amssymb,amsfonts}
\usepackage{amsthm}
\usepackage{mathrsfs}
\usepackage[title]{appendix}
\usepackage{xcolor}
\usepackage{textcomp}
\usepackage{manyfoot}
\usepackage{booktabs}
\usepackage{algorithm}
\usepackage{algorithmicx}
\usepackage{algpseudocode}
\usepackage{listings}

\theoremstyle{thmstyleone}

\theoremstyle{thmstyletwo}

\theoremstyle{thmstylethree}

\newcommand{\la}{\left\langle}
\newcommand{\ra}{\right\rangle}
\newcommand{\lp}{\left(}
\newcommand{\rp}{\right)}
\newcommand{\e}{\mathrm e}

\begin{document}

\title[Synchronization transitions in Kuramoto oscillators with mixed interactions]{Synchronization transitions in Kuramoto oscillators with mixed interactions}

\author[1]{\fnm{Abhishek} \sur{Sharma}}\email{abhishek1intel@gmail.com}

\author[1]{\fnm{Ayushi} \sur{Suman}}\email{ayushishuman@gmail.com}

\author[2,3]{\fnm{Mehernaz} \sur{Anvari}}\email{mehrnaz.anvari@scai.fraunhofer.de}

\author[4]{\fnm{Manish Dev} \sur{Shrimali}}\email{shrimali@curaj.ac.in}

\author[5,6,7]{\fnm{Charo I.} \sur{del Genio}}\email{the.paraw@gmail.com}

\author*[1]{\fnm{Sarika} \sur{Jalan}}\email{sarikajalan9@gmail.com}

\affil[1]{\orgdiv{Complex Systems Lab, Department of Physics}, \orgname{Indian Institute of Technology Indore}, \orgaddress{\street{Khandwa Road}, \city{Indore}, \postcode{453552}, \country{India}}}

\affil[2]{\orgname{Fraunhofer Institute for Algorithms and Scientific Computing}, \orgaddress{\street{Schloss Birlinghoven}, \city{Sankt Augustin}, \postcode{53757}, \country{Germany}}}

\affil[3]{\orgname{Potsdam Institute for Climate Impact Research}, \orgaddress{\street{Telegrafenberg}, \city{Potsdam}, \postcode{14473}, \country{Germany}}}

\affil[4]{\orgdiv{Department of Physics}, \orgname{Central University of Rajasthan}, \orgaddress{\street{NH-8 Bandar Sindri}, \city{Ajmer}, \postcode{305817}, \country{India}}}

\affil[5]{\orgdiv{Institute of Smart Agriculture for Safe and Functional Foods and Supplements}, \orgname{Trakia University}, \orgaddress{\street{Studentski Grad}, \city{Stara Zagora}, \postcode{6000}, \country{Bulgaria}}}

\affil[6]{\orgdiv{Institute of Interdisciplinary Intelligent Science}, \orgname{Ningbo University of Technology}, \orgaddress{\street{201 Fenghua Road}, \city{Ningbo}, \postcode{315211}, \country{China}}}

\affil[7]{\orgdiv{School of Mathematics}, \orgname{North University of China}, \orgaddress{\street{3 College Road}, \city{Taiyuan}, \postcode{030051}, \country{China}}}

\abstract{The Kuramoto model is a paradigmatic system
for emergent synchronization phenomena, and significant
efforts have been made to investigate its transition
from incoherent to coherent dynamics. However, previous
studies have almost exclusively assumed global coupling
between the oscillators, even though many real-world
systems are organized in specific network structures.
Here, we fill this gap by examining finite-size effects
in Kuramoto oscillator ensembles on networks with mixed-order
interactions. Using numerical simulations, we show how
parameters such as the size of the system, the average
degree of the nodes and the strength of the interactions
alter the behaviour of the transition to synchronization
with respect to the known theoretical results that are
derived for globally coupled infinite systems.}


\maketitle

\section{Introduction}
Synchronization in dynamical systems is an emergent phenomenon
in which the trajectories of a population of coupled oscillators
reach some level of macroscopic coherence induced by their interactions.
Since coherent collective behaviour is commonly necessary for
the correct functioning of complex systems, the study of synchronization
has received considerable attention in a broad range of topics
and disciplines~\cite{Pik03,Are08,del15,del16,Boc18,del22,Jaf24}.
A paradigmatic framework for its investigation is provided by
the Kuramoto model, in which oscillators interact via a pairwise
sinusoidal coupling that can induce spontaneous phase locking
once its strength exceeds a critical threshold~\cite{Kur75}. Owing
to its relative simplicity and analytical tractability, the Kuramoto
model and its variants have become very popular in the study of
a wide range of real-world complex systems~\cite{Erm91,Sep04,Chi08,Mar09,Bre10,Hon11,Zhu15,OKe17,Osa17}.

Recently, however, there has been a growing scientific interest
in studying the phenomenonology of systems in which the interactions
between elements are of a higher order than just pairwise ones~\cite{Kac1,Kac2,Boc23,Ma25,del25,del26}.
The inclusion of higher-order interactions in models of complex
systems has thus raised new questions about their impact on collective
behavior, especially when combined with traditional pairwise couplings.
Understanding these combined effects is crucial, as both pairwise
and higher-order interactions coexist in real-world systems, and
their synergy can give rise to highly non-trivial effects~\cite{Par25,Sem25,Kov26,Guo26}.
In the specific case of globally coupled Kuramoto oscillators,
the presence of higher-order interactions is known to cause a histeresis
loop in the synchronization transition, which is therefore a first-order
one~\cite{Dut23,Raj23,Bis24,Pal24,Lou25}, in contrast with the continuous
transition that occurs when only pairwise interactions exist.

Additionally, most investigations on Kuramoto oscillators
have focused on large ensembles where analytical results
are typically derived in the thermodynamic limit. However,
numerous physical and biological systems have small sizes
that are not well approximated by an infinite system. It
becomes thus essential to study how finite-size fluctuations
can affect the emergent collective dynamics of commonly used
models. This is a particularly relevant endeavour since their
effects can be highly non-trivial~\cite{Dai87,Dai90,Dai96,Sum24}.
For example, in oscillators with only higher-order coupling,
synchronization transitions arise only in finite systems,
and disappear in the thermodynamic limit~\cite{Kom15,Pet18}.
However, all the studies on finite-size effects have been
restricted so far to all-to-all topologies.

Here, we carry out a numerical study of the dynamics
of a system consisting of a finite number of oscillators
connected in a traditional, pairwise, Erd\H{o}s-R\'enyi~(ER)
random graph with the addition of higher-order global
connections, revealing the effects that system size,
density and coupling strengths have on the synchronization
transition, and highlighting the differences with globally
coupled systems in the thermodynamic limit. We have
introduced mixed interactions to understand the finite
size effect in a system with network topology and distinction
between network and globally connected system.

\section{Results}
\subsection{The model}
Consider a system of $N$~interacting phase oscillators,
each with characteristic frequency~$\omega_i$, interacting
via two different couplings, namely a pairwise one with
strength~$K_1$ and a triadic one with strength~$K_2$. The
pairwise interactions occur over the edges of an ER random
network with connection probability~$p$, yielding an average
degree $\la k\ra = p\lp N-1\rp$, whereas the higher-order
interactions are all-to-all. The dynamics of the system
is described by
\begin{equation}\label{model}
 \dot \theta_i = \omega_i + \frac{K_1 }{\la k\ra} \sum_{j=1}^{N} A_{ij}\sin(\theta_j - \theta_i ) +  \frac{K_2}{N^2} \sum_{j,l=1} ^N \sin (2\theta_j -\theta_l - \theta_i)\:,
\end{equation}
where $A_{ij}$ is the~$(i,j)$ element of the adjacency matrix
of the ER network.

\begin{figure}[t]
 \centering
 \includegraphics[width=0.9\textwidth]{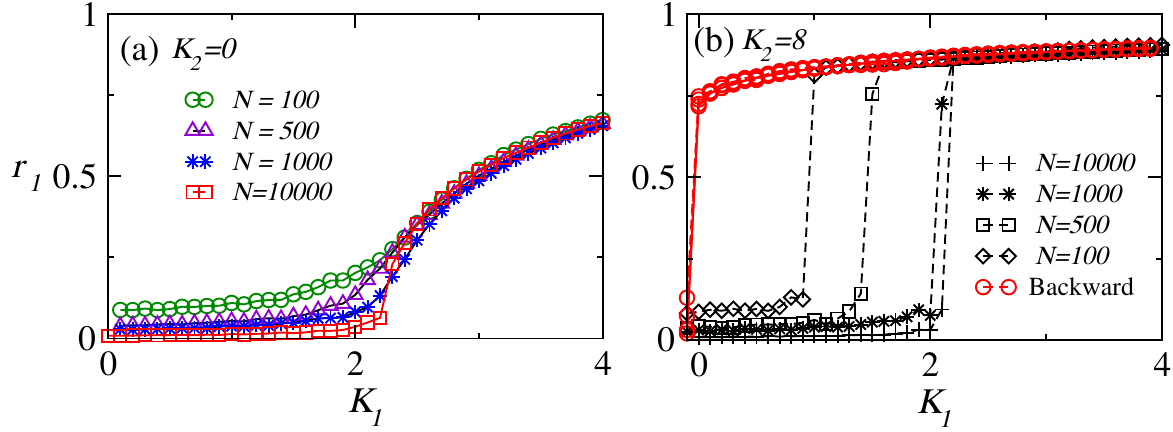}
 \caption{\textbf{Higher-order interactions change the
 character of the synchronization transition.} (a) The
 global order parameter changes continuously with the
 pairwise coupling strength in the absence of higher-order
 interactions. (b) When the higher-order coupling strength
 is set to $K_2=8$, the synchronization transition becomes
 first-order, with the appearance of a hysteresis loop.
 Also, the forward transition point changes with system
 size, whereas the backwards one remains constant. In
 all cases, the average degree of the network is fixed
 at~20.}\label{Kvsr_ER}
\end{figure}
To study the system, we measure the local order parameter~$z^{(i)}$,
which quantifies the degree of local synchrony between the $i$th~oscillator
and its neighbours, and the Kuramoto-Daido global order parameters~$z_q$,
which characterize the extent of higher-harmonics synchronization of
the entire population:
\begin{alignat}{4}
 z^{(i)} &= \frac{1}{\la k\ra}\sum_{j=1}^{N} A_{ij}\e^{\iota \theta_j} &=& r_1^{(i)} \e^{\iota \psi_1^i}\:,\\
 z_q &= \frac{1}{N} \sum_{j=1}^N \e^{\iota q \theta_j} &=& r_q \e^{\iota \psi_q} \:.
\end{alignat}
In the equations above, the magnitude of the local order parameter, $r_1^{(i)}$,
is the fraction of oscillators that are entrained to oscillator~$i$,
and $\psi_1^{(i)}$ is the local mean phase. The Kuramoto-Daido
order parameters, instead, correspond to the Fourier modes of
the probability density of phases. Thus, their magnitudes~$r_q$
measure the extent by which the system splits into $q$~synchronized
clusters with phase difference~$\frac{2\pi}{q}$ between them.
Eq.~\ref{model} can then be rewritten as
\begin{equation}
 \dot \theta_i = \omega_i + K_1 r_1^{(i)} \sin(\psi_1^i - \theta_i) + K_2 r_2 r_1 \sin(\psi_2 - \psi_1-\theta_i)\:.\label{Meanfield}
\end{equation}

\begin{figure}[t]
 \centering
 \includegraphics[width=0.9\textwidth]{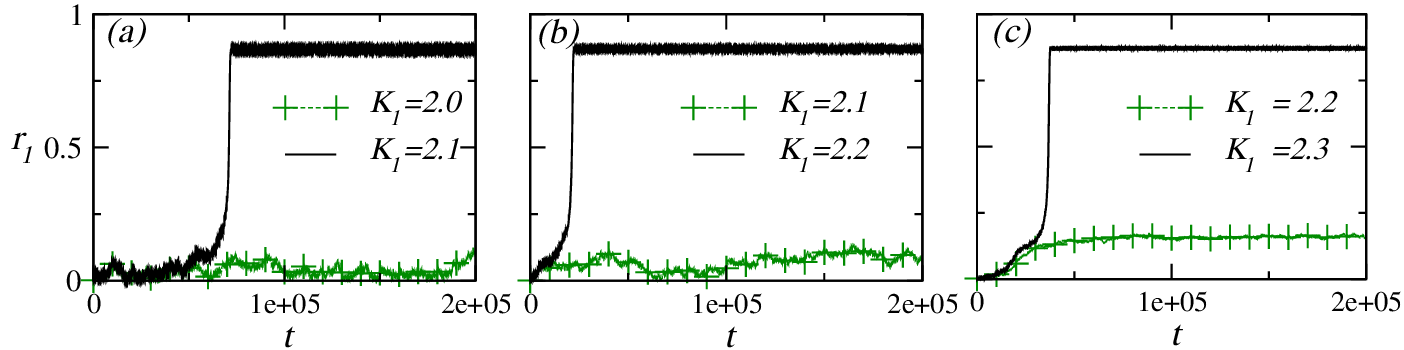}
 \caption{\textbf{The transition point in random networks is
 not bounded by~2}. The time evolution of the magnitude of the
 Kuramoto order parameter~$r_1$ for $N=10^4$~(a), $N=2\times10^4$~(b),
 and $N=10^5$~(c) shows that the abrupt transition to synchronization
 happens for values of the coupling strength~$K_1$ that are
 greater than the theoretical limit of~2 that is valid for complete
 graphs. In all cases, the higher-order coupling was $K_2=8$,
 and the initial conditions were completely  random, corresponding
 to $r_1=0$.}\label{rvst}
\end{figure}
For the sake of completeness, we note that the equation above,
for a globally coupled system and in the thermodynamic limit,
admits a low-dimensional solution using the Ott-Antonsen Ansatz~\cite{Ott08}
that provides the time evolution for~$r_1$:
\begin{equation}\label{all2allanalytics}
 \dot r_1 = -r_1 + \frac{K_1}{2}\lp r_1-r_1^3\rp+ \frac{K_2}{2} \lp r_1^3-r_1^5\rp\:.
\end{equation}

\subsection{Characterization of the synchronization transition}
To investigate finite-size effects on the behaviour of the system,
we solve Eq.~\ref{Meanfield} numerically, varying the system size
from~$N=10^2$ to~$N=10^4$, while keeping the average degree fixed
at~20, to ensure that the network remains well above the connectivity
threshold of an ER~graph. Also, we sample the characteristic frequencies
from a Lorentzian distribution $\mathcal{L}(0,1)=\frac{1}{\pi}\frac{1}{x^2+1}$,
for which the critical coupling for the transition to synchronization
is known to have a theoretical upper bound of~2~\cite{Jal22}.

As illustrated in Fig.~\ref{Kvsr_ER}(a), in the absence
higher-order interactions, the transition from incoherent
dynamics to a synchronized state is continuous, and the
onset of synchronization is independent of system size,
apart from scaling corrections in the vicinity of the critical
point~\cite{Hon15}. Upon inclusion of higher-order coupling,
the transition becomes abrupt and it is accompanied by
a bistable region, as shown in Fig.~\ref{Kvsr_ER}(b).
Moreover, a clear dependence emerges of the forward transition
point with system size.

\begin{figure}[t] 
 \centering
\begingroup
\begin{tabular}{c c c}
 \includegraphics[width=0.333\textwidth]{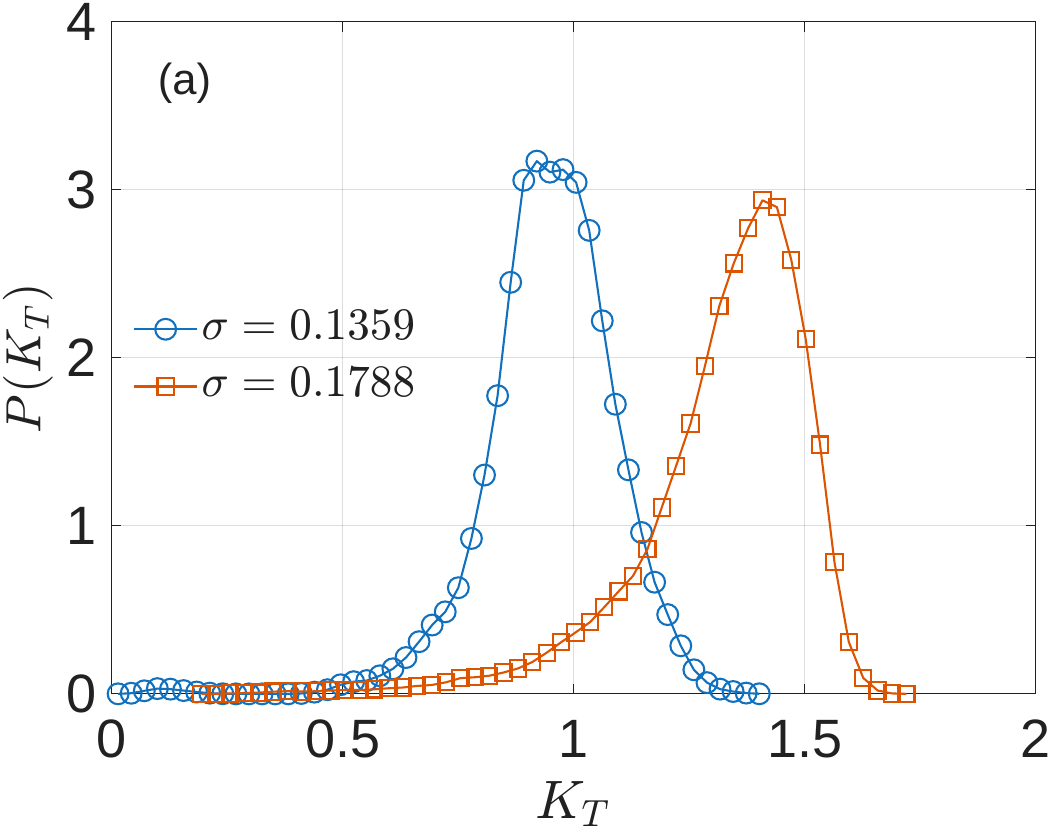}
 \includegraphics[width=0.333\textwidth]{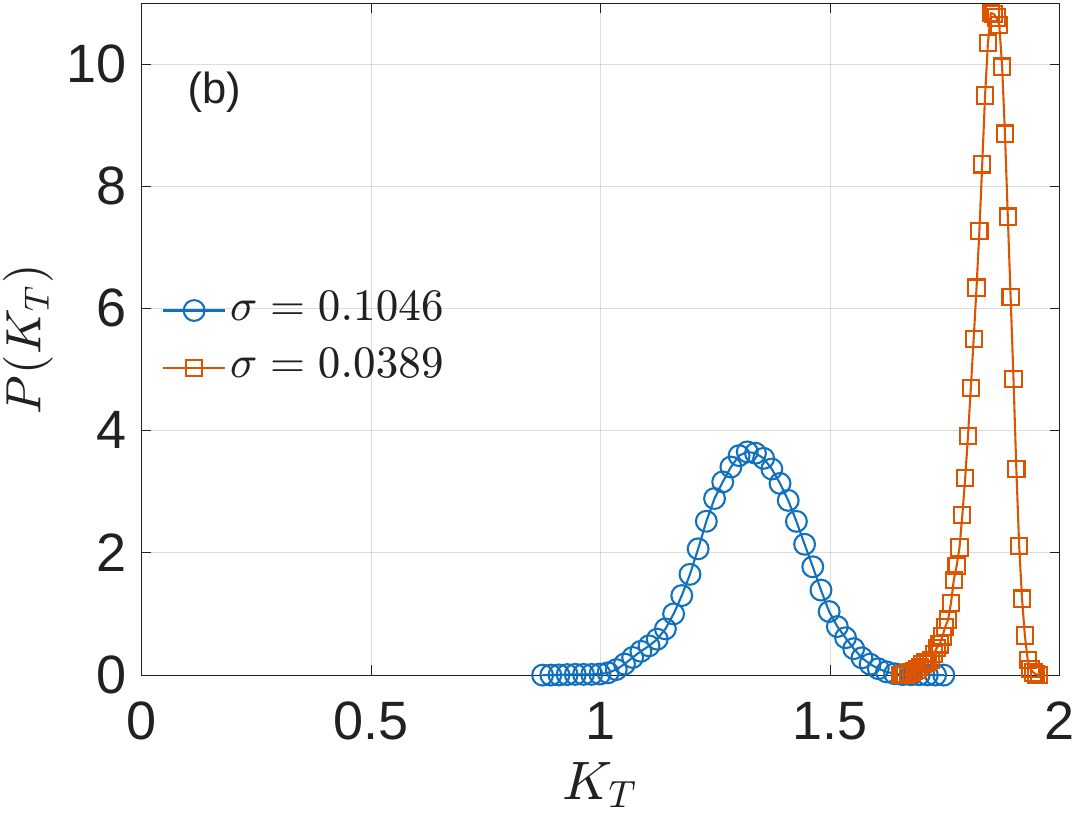}
 \includegraphics[width=0.333\textwidth]{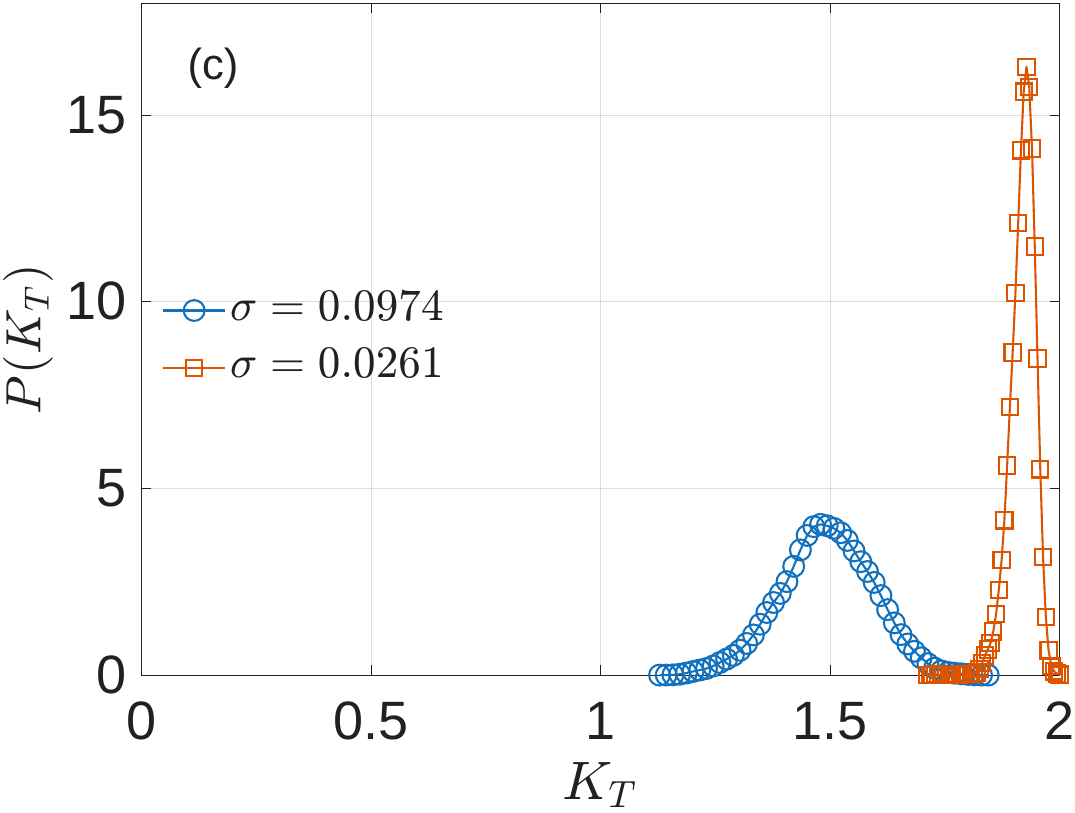} 
 &\\
 \includegraphics[width=0.333\textwidth]{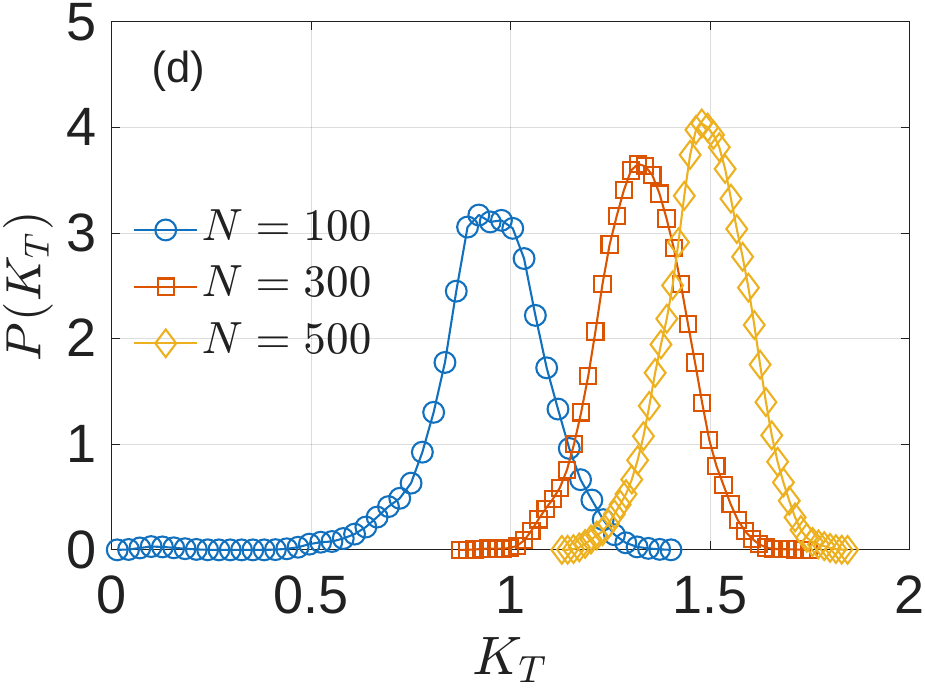}
 \includegraphics[width=0.333\textwidth]{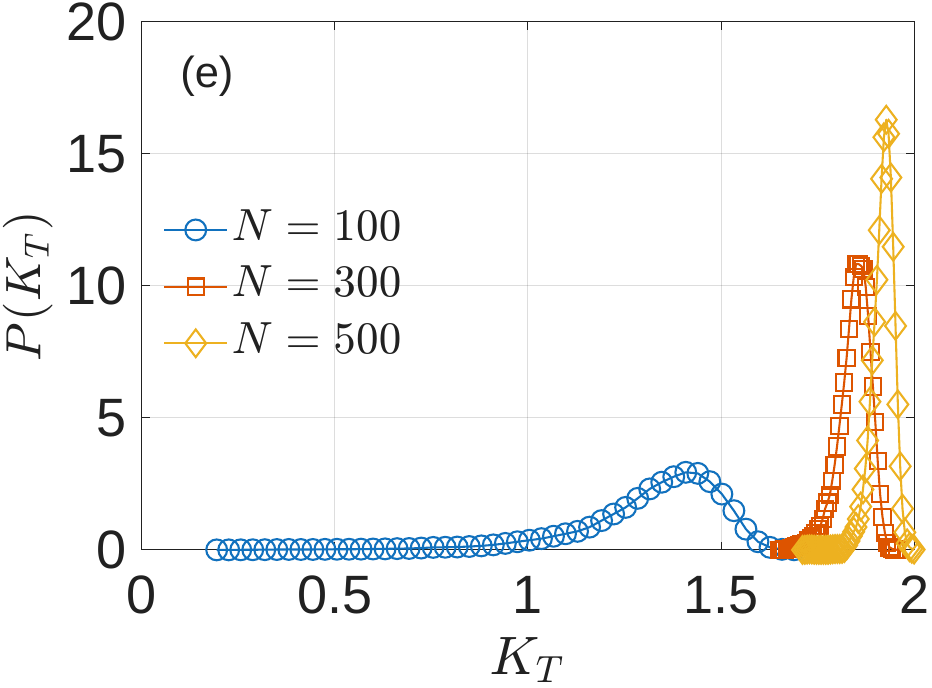}
 \end{tabular}
 \endgroup
 \caption{\textbf{The mean transition coupling strength
 increases with system size.} (a)--(c) The probability
 distributions of the transition points~$K_T$ shift
 towards higher values and become narrower as the
 size of the ER~network increases from~100 to~300
 and~500. The same qualitative behaviour is found
 for complete graphs. (d)--(e) The same data are
 plotted by network type.}\label{CompareERandGlobal}
\end{figure}
Note that, as we mentioned above, in a globally coupled system,
the forward critical coupling approaches~2 in the thermodynamic
limit. In contrast, here we find that, with mixed-order interactions,
the transition point exceeds this theoretical maximum for $N=10^4$.
To better show this, in Fig.~\ref{rvst} we report the time evolution
of $r_1$ for large but finite system sizes. The magnitude of
the Kuramoto order parameter shows that the system undergoes an
abrupt transition to the synchronized state at distinct coupling
strengths that are larger than~2, confirming that, in contrast
to what happens on complete graphs, the transition point in a random
network is not bounded by~2.

To better characterize the effect of network size
on the synchronization transition in the presence
of mixed couplings, we simulated the system over
3000~realizations, each corresponding to a different
network structure and with different initial conditions,
and recorded the transition point~$K_T$ for each.
$K_T$ denotes the critical coupling strength
at which the system jumps from a stable incoherent
state and undergoes a transition to a stable coherent
state; this value is referred to as the transition
point. For each realization, characterized by distinct
initial conditions and network topologies, we recorded
the coupling value at the transition point and subsequently
constructed the probability distribution of~$K_T$.
The distributions of transition points, reported
in Fig.~\ref{CompareERandGlobal}, is unimodal, with
the peak that shifts toward larger coupling strengths
for increasing system sizes. The evolution of the distribution
is more rapid for globally connected systems than
it is for random networks of the same size, even
though the qualitative behaviour is the same. Also,
in general, in ER~networks, the peak is found at
lower couplings than those at which it occurs in
complete graphs.

\subsection{Estimation of the unstable fixed points}
\begin{figure}[t]
 \centering
 \begingroup
 \begin{tabular}{c c c}
 \includegraphics[width=0.333\textwidth]{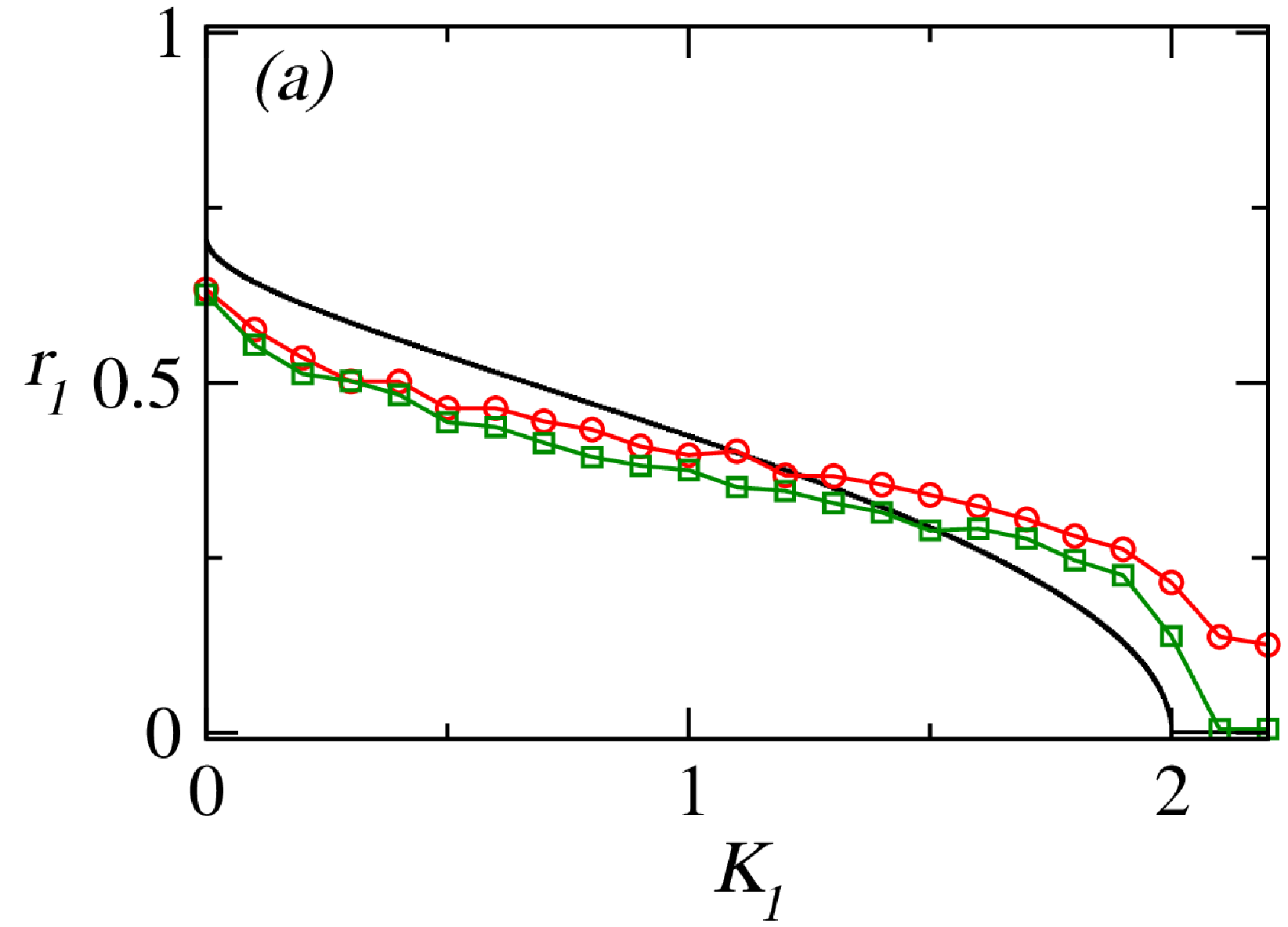}
 \includegraphics[width=0.333\textwidth]{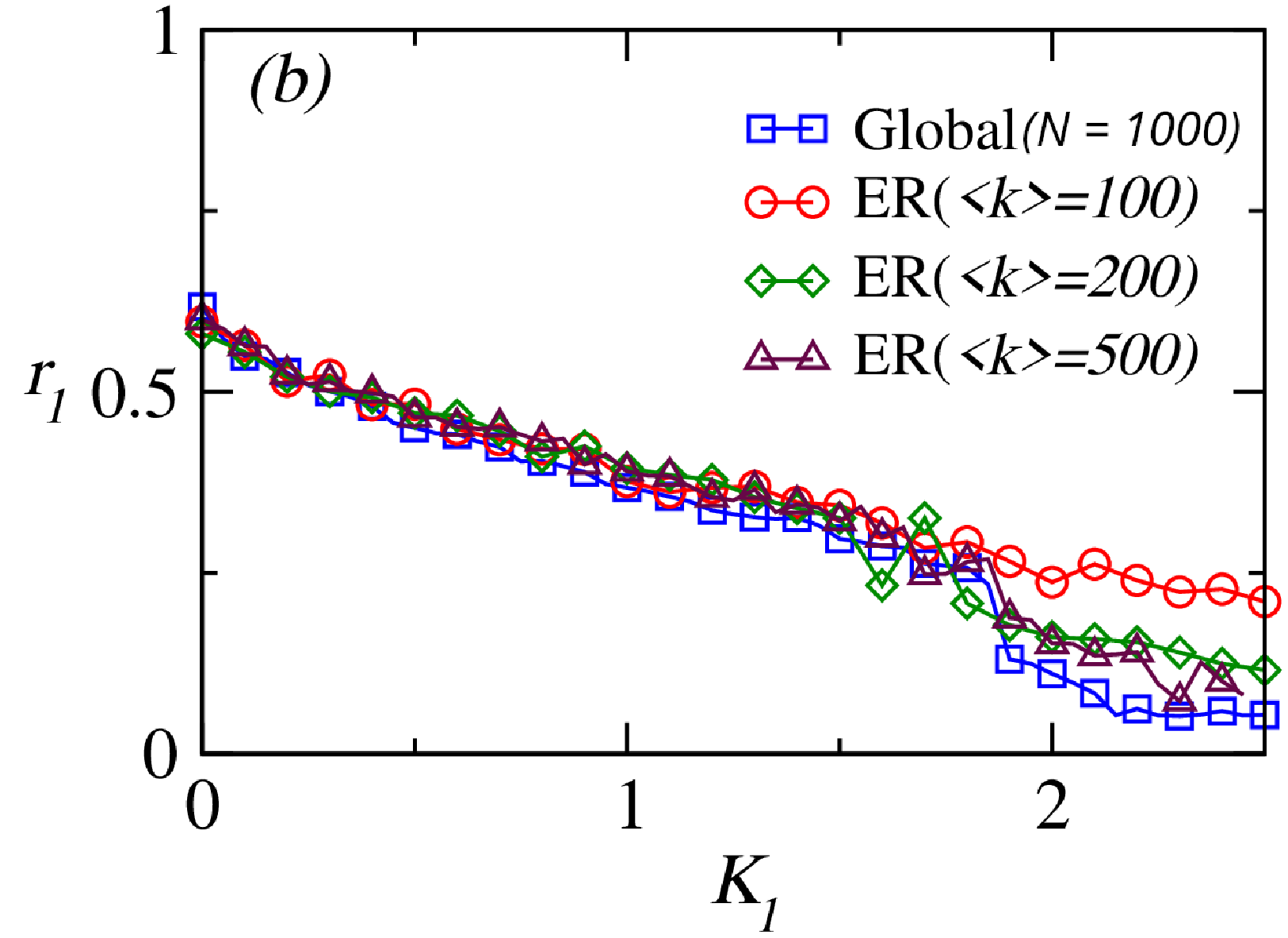}
 \includegraphics[width=0.333\textwidth]{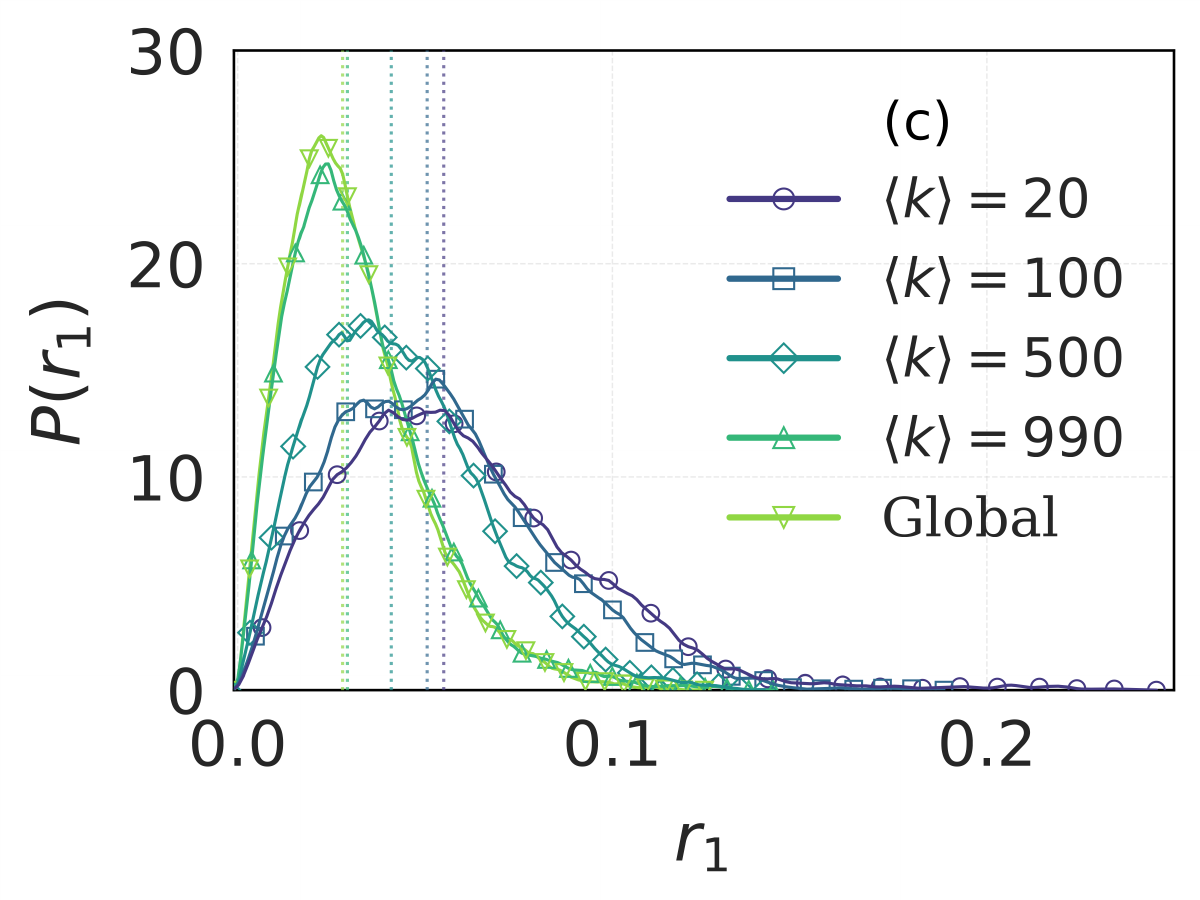}
 \end{tabular}
 \endgroup
 \caption{\textbf{Crossover of the sign of finite size effects
 on unstable fixed points.} (a) For low values of~$K_1$, the locations of the unstable
 fixed point for an ER~network with $10^4$~nodes and average
 degree~20 (red), and for a complete graph of the same size
 (green) are lower than those for an infinite complete graph
 (black), obtained by solving Eq.~\ref{all2allanalytics}. However,
 their shift is in the opposite direction for high values of~$K_1$.
 (b) The unstable fixed points for ER~networks with 1000~nodes
 approach the behaviour of a complete graph in the thermodynamic
 limit as the average degree is increased. (c)$P(r_1)$ versus $r_1$. Probability distribution of $r_1$ for the incoherent state with $N=1000$, $K_2=8$, and $K_1=1.5$, computed over a time window of $200000$.}\label{unstable fixed points}
\end{figure}
The trend of the distributions of transition points
reported in the previous section can be explained by
means of unstable fixed points and fluctuations in
order parameter. However, a theoretical estimate of
the unstable state can only be provided for the case
of globally coupled oscillators in the thermodynamic
limit. Thus, we use a numerical approach to identify
them in the case of finite systems with random-network
topology.

Specifically, for each fixed value of the pairwise coupling
strength~$K_1$, we initialize the phases~$\theta_i$ to generate
a set of initial order parameter values~$r_1$ in the range
$[0.01, 0.7]$. Each realization of the system is subsequently
evolved to obtain an updated order parameter value~$r_1'$.
In our simulations, the resulting value of the order parameter
is always smaller than~$0.1$ or greater than~$0.6$. Thus, we
adopt a discriminating criterion based on this observation:
if $r_1' > 0.6$, the corresponding initial conditions are classified
as belonging to the coherent basin, whereas if $r_1' < 0.1$,
they are classified as belonging to the incoherent one. This
allows us to determine, for each value of~$K_1$, the largest~$r_1'$
associated with a coherent state and the smallest~$r_1'$
associated with an incoherent state. The unstable fixed point
is then the arithmetic mean of these two extreme values.

The results, illustrated in Fig.~\ref{unstable fixed points}(a)
for $N=10^4$ in the case of an ER~network with $\la k\ra=20$, as
well as in the case of a complete graph, show a behaviour of the
unstable fixed point that is similar to that of a globally connected
infinite network obtained by solving Eq.~\ref{all2allanalytics}.
The finite size of the system, however, causes it to shift towards
lower values of~$r_1$ for low~$K_1$, and towards higher values
of~$r_1$ for high~$K_1$, with a crossover at $K_1\approx 1.25$.
In fact, the unstable fixed point reaches~0 only at $K_1>2$ even
for the complete graph.

Furthermore, for higher values of the coupling strength,
the distance of the unstable branch from $r_1 = 0$ decreases
with the average degree, as shown in Fig.~\ref{unstable fixed points}(b).
This is consistent with the previous observations, and it
indicates that as the average degree approaches the value
it has in fully connected graphs, the unstable branch moves closer
to~0 as $K_1$ approaches~2.

The effects of the average degree are measurable
also for low coupling strengths. In fact, the distribution
of values of~$r_1$ for networks of size~1000 and
$K_1=1.5$, shown in Fig.~\ref{unstable fixed points}(c),
indicate that the average value of~$r_1$ decreases
with the average degree. This complements the results
shown in Fig.~\ref{Kvsr_ER}(a), from which one can
notice that the baseline of the order parameter
in the incoherent state scales like $\frac{1}{\sqrt{N}}$,
consistently with known theoretical treatments~\cite{Str00}.
Additionally, both the mean and the standard deviation
of the distribution decrease with the average degree,
implying that fluctuations in the order parameter,
which are relevant for the transition to synchronization,
are smaller in complete graphs than they are in
ER~networks.

\subsection{Effects of density and higher-order coupling}
\begin{figure}[t]
 \centering
 \begingroup
 \begin{tabular}{c c c} 
 \includegraphics[width=0.333\textwidth]{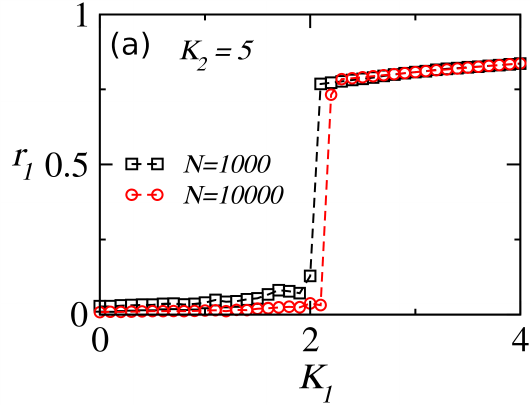}
 \includegraphics[width=0.33\textwidth]{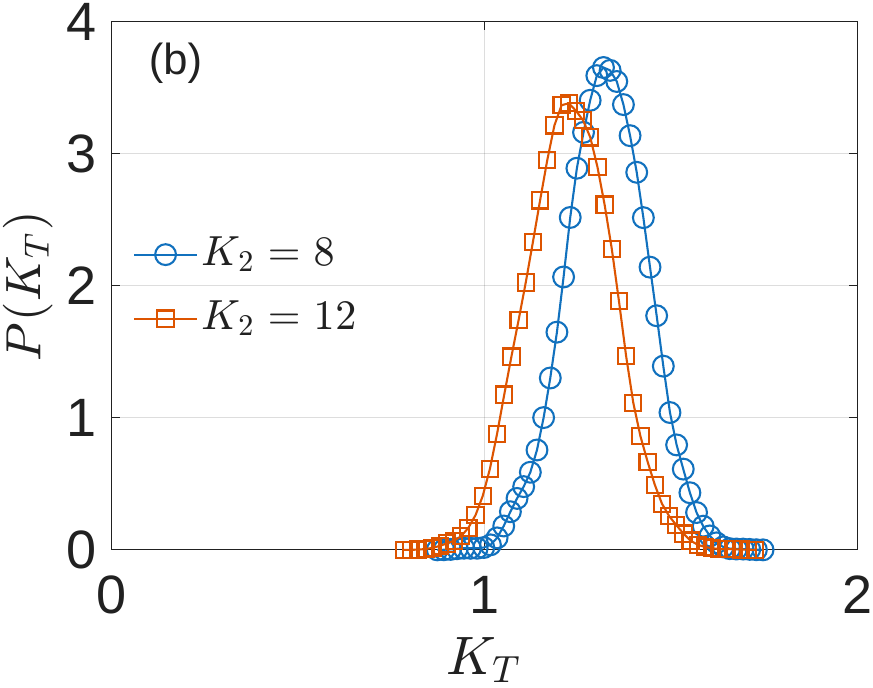}
 \includegraphics[width=0.33\textwidth]{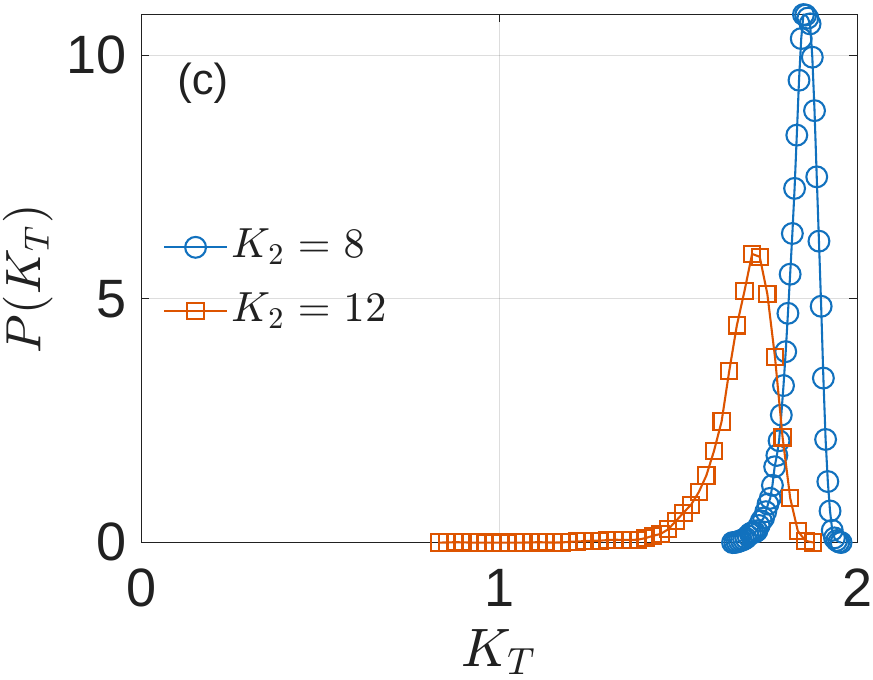}
  \end{tabular}
\endgroup
 \caption{\textbf{System size and higher-order coupling
 have opposite effects on the transition point.} (a) For
 a fixed value of~$K_2$, increasing the number of nodes
 in the network shifts the transition point to higher
 values of~$K_1$. (b)--(c) Increasinng the higher-order
 coupling shifts the distribution of transition points
 towards lower values and increases its width, with
 the effect being more dramatic on fully connected networks~(c)
 than on ER~random graphs~(b). The results are averaged
 over~3000 independent realizations, and the system
 size is~300 in panels~(b) and~(c).}\label{P(Kc)K2}
\end{figure}
Further simulations reveal that system size and higher-order
coupling have opposite effects on the transition point. In fact,
the value of~$K_1$ at which the synchronization transition happens
increases with the number of oscillators, as shown in Fig.~\ref{P(Kc)K2}(a).
Note that, while the results reported in the figure are for
$K_2=5$, the same behavour occurs regardless of the strenght
of the higher-order coupling.

Conversely, an increase in~$K_2$ for a fixed
system size leads to a leftward shift in the
distribution of~$K_T$ regardless of the structure
of connections, as illustrated in Fig.~\ref{P(Kc)K2}(b)--(c).
This is caused by a change in the location of
the unstable fixed points, which depends on
the value of~$K_2$. In turn, the difference
in this change between ER~networks and complete
graphs causes the effect of~$K_2$ to be stronger
for fully connected networks than for random
ones.

\begin{figure}[t]
 \centering
 \includegraphics[width=0.4\textwidth]{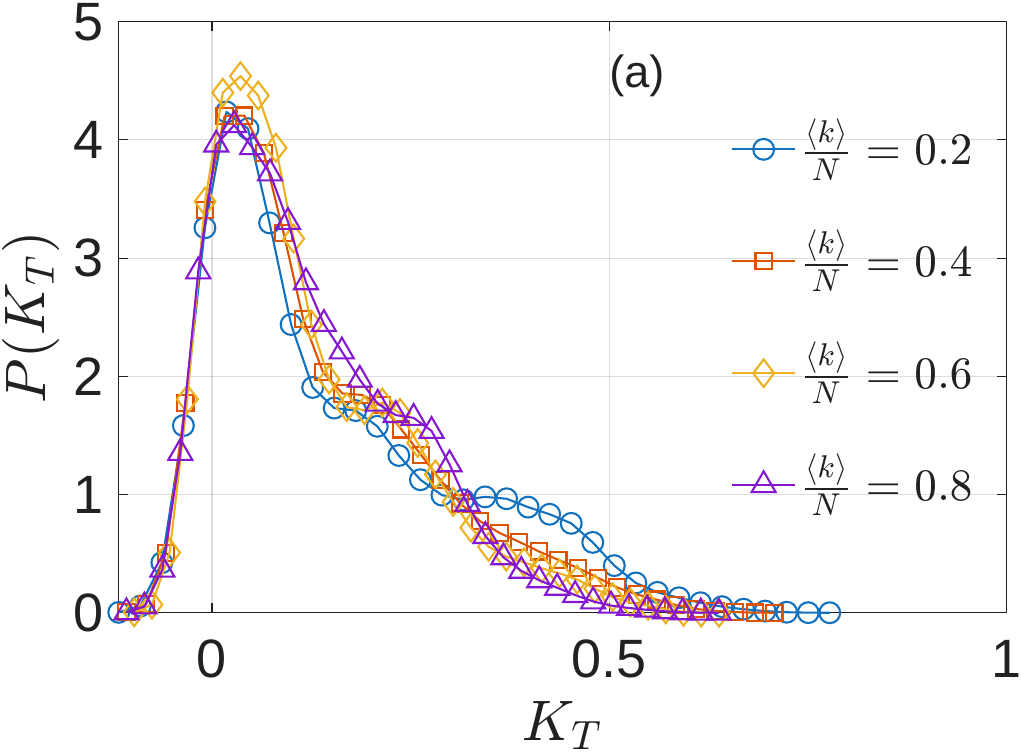}
 \caption{\textbf{Density has only a minor effect on
 the distribution of transition points.} For a system
 of fixed size ($N=50$) and fixed higher-order coupling
 ($K_2=8$), increasing the density does not shift the
 peak of the distribution of~$K_T$. The only noticeable
 effect is a smoothening of the distribution for higher
 values of~$K_1$, due to the smaller relative importance
 of fluctuations.}\label{picn50p004}
\end{figure}

Additionally, we find that the density of the network
has only a minor effect compared to the network size,
especially for small systems. In fact, we have already
shown that, as the average degree of a network increases,
the baseline value of the order parameter in the incoherent
state decreases. However, the ultimate influence of this
change on the distribution of the forward transition
point is fairly small, as shown in Fig.~\ref{picn50p004},
with no noticeable shift in its peak. In contrast, increasing~$N$
for a fixed density clearly shifts the distribution towards
higher values of~$K_T$, reducing its standard deviation
and making the peak sharper, as already illustrated in
Fig.~\ref{CompareERandGlobal}(d).

\section{Conclusion}
In summary, we carried out an extensive numerical investigation
of the synchronization transition of Kuramoto oscillators in systems
with mixed-order coupling. Specifically, we imposed pairwise interactions
along the edges of Erd\H{o}s-R\'enyi random graphs, together with
all-to-all three-body interactions. Also, crucially, we focused
on systems whose sizes are small enough that finite-size effects
play a significant role.

In such systems, fluctuations in the stationary value
of the order parameter are strongly influenced by system
size, so that, given enough time, they can drive the
system across the basin boundary of the unstable
fixed point and induce a transition from incoherence
to coherence. As the system size increases, fluctuations
become weaker, and the system remains confined below
the unstable branch, eventually preventing a transition
in the thermodynamic limit.

Indeed, our simulations demonstrate that the forward synchronization
transition is strongly affected by the fluctuations of the order parameter,
which induce a size-dependent shift in the coupling strength at which
the transition happens. For relatively small ER~networks, the effective
transition point is lower than that of the corresponding globally coupled
system. In contrast, for large enough system size, globally coupled
systems always synchronize at $K_1\rightarrow 2$, whereas ER~networks
require a larger coupling strength. We describe this result by numerically
identifying the unstable fixed points of the system, and showing how
for low pairwise couplings, their value is lower than that of complete
graphs, crossing over to higher values for high coupling strengths.

Finally, we examined the distribution of transition thresholds
across a large ensemble of network realizations and initial conditions.
This distribution exhibits a well-defined peak, and its width
decreases with system size, reflecting the gradual suppression
of fluctuations.

As most real-world systems with random network topologies possess finite sizes that cannot be accurately represented by the thermodynamic limit, our results underline
the importance of accounting for finite-size effects when
studying and interpreting synchronization transitions with network topology.

\backmatter

\bmhead{Acknowledgements}
Abhishek Sharma acknowledges UGC for a Senior Research Fellowship.
Ayushi Suman acknowledges CSIR award SRF-NET2021/11014.
Mehernaz Anvari acknowledges IGSTC WISER award with Grant No.~11-20286-3180-00001.
Charo del~Genio acknowledges funding from the Bulgarian Ministry of Education and Science, under Project No.~BG-RRP-2.004-0006-C02.
Sarika Jalan acknowledges SERB POWER Grant No.~SPF/2021/000136.

\section*{Declarations}
\subsection*{Competing interests}
The authors declare no competing interests.

\subsection*{Author contribution}
SJ and CIDG developed the research concept and supervised the study.
AS conducted the numerical simulations. All authors contributed to the
analysis and interpretation of the results. AS, MS, CDG and SJ prepared
the manuscript. All authors approved the manuscript.
\bibliography{Kuramix}

\end{document}